\documentclass[12pt]{article}
\usepackage{epsfig,epsf,amssymb,amsmath,latexsym,bbm}
\title{Positive Lyapunov exponents and localization bounds \\
for strongly mixing potentials}

\author{Christian Sadel, Hermann Schulz-Baldes
\\
\\
{\small Mathematisches Institut, Universit\"at
Erlangen-N\"urnberg, Germany}
}

\date{ }

\newtheorem{theo}{Theorem}

\newtheorem{proposi}{Proposition}
\newtheorem{lemma}{Lemma}

\newtheorem{rem}{Remark}

\newcommand{\CC}{{\mathbb C}}
\newcommand{\NN}{{\mathbb N}}
\newcommand{\RR}{{\mathbb R}}

\newcommand{\ZZ}{{\mathbb Z}}

\newcommand{\VM}{{\mathbb V}}

\newcommand{\Pp}{{\cal P}}
\newcommand{\PP}{{\bf P}}

\newcommand{\EE}{{\bf E}}

\newcommand{\Gg}{{\cal G}}

\newcommand{\Ss}{{\cal S}}
\newcommand{\Oo}{{\cal O}}

\newcommand{\Tt}{{\cal T}}

\newcommand{\Ll}{{\cal L}}
\newcommand{\Qq}{{\cal Q}}

\def\esssup{\mathop{\rm ess\,sup}}

\newcommand{\one}{{\bf 1}}

\newcommand{\comm}[1]{}
\newcommand{\comment}[1]{}

\newcommand{\Var}{{\rm Var}}

\begin{document}

\maketitle

\begin{abstract}
For a one-dimensional discrete Schr\"odinger operator
with a weakly coupled potential given by
a strongly mixing dynamical system with power law decay of correlations, 
we derive for
all energies including the band edges and the band center
a perturbative formula for the Lyapunov exponent.  
Under adequate hypothesis, this shows that the Lyapunov exponent is positive on
the whole spectrum. This 
in turn implies that the Hausdorff dimension of the
spectral measure is zero and that the associated quantum dynamics grows at most
logarithmically in time.
\end{abstract}


\section{Introduction}

Let $\Sigma$ be a topological space and $\Omega=\Sigma^{\ZZ}$ the associated
Tychonov product space. Furthermore let $\PP$ be a propability measure 
on $\Omega$ which is invariant and ergodic w.r.t. the left shift
$S:\Omega\to\Omega$. Now given a measurable real-valued 
function $V$ on $\Omega$ and a coupling constant $\lambda> 0$, one can
associate an ergodic family of Jacobi matrices
$(H_{\lambda,\omega})_{\omega\in\Omega}$  (also called 
discrete Schr\"odinger operators)
each acting on $\ell^2(\ZZ)$:
\begin{equation}
\label{eq-Jacobi}
H_{\lambda,\omega}\, |n\rangle
\;=\;
|n+1\rangle \,+\, \lambda\, V(S^n \omega)\,|n\rangle \,+\, |n-1\rangle\;,
\end{equation}
where $|n\rangle$ is the Dirac notation for the state in
$\ell^2(\ZZ)$ localized at site $n\in\ZZ$. If 
$\PP={\bf p}^{\otimes\ZZ}$ is a product measure  of a compactly supported
probability measure ${\bf p}$ on $\Sigma$ so that the random variables
of the sequence
$(V(S^n \omega))_{n\in\ZZ}$ of potential values 
are independent, the model exhibits
so-called Anderson localization, namely the spectrum of 
$H_{\lambda,\omega}$ is $\PP$-almost surely pure-point with exponentially
localized eigenstates \cite{PF} and the induced quantum dynamics is
bounded in time (in the precise sense given below).
The question considered in this
work (and many others, see the reviews \cite{Jit,Dam} 
and references therein) concerns the spectral properties 
as well as the quantum dynamics in situations where $\PP$ is not a product
measure so that the random variables $(V(S^n \omega))_{n\in\ZZ}$ are
correlated. This situation typically 
arises when the dynamical system $(\Omega,S,\PP)$ 
is the symbolic dynamics associated to a (possibly weakly) hyperbolic 
discrete time dynamics; then $\Sigma$ is the Markov partition. 
If now the correlations of the potential decay sufficiently fast, 
then one expects that the model is still in the
regime of Anderson localization. Here we complement on the prior work 
\cite{CS,BS} and prove that this holds
at least in a weak sense when the correlations satisfy
a power law decay. 

\vspace{.2cm}

The proof of localization for these models is
based on the positivity of the Lyapunov exponent. This positivity
can either be
established by Kotani theory \cite{Dam}, a version of Furstenberg's theorem
for correlated random matrices (work by Avila and Damanik cited in 
\cite{Dam}) or by a perturbative calculation (for
small $\lambda$) of the Lyapunov exponent. This latter calculation 
was first done by
Chulaevski and Spencer \cite{CS} by carrying over the argument of Thouless
\cite{Tho}, in a version given by Pastur and Figotin \cite{PF}, to the case of
correlated potential values. The resulting formula is recalled in
Section~\ref{sec-results}. Based on this result, Bourgain and
Schlag then proved localization \cite{BS}.
The only flaw left is that in \cite{CS} (and actually already in \cite{PF}) 
not all energies could be dealt with, but the band center and
the band edges were spared out. Here we show how
the techniques of our prior works on anomalies and band edges \cite{S,SS}
combine with those of \cite{CS} to rigorously control the perturbation theory
for the Lyapunov exponent also at these energies. Instead of repeating the
rather complicated proofs of \cite{BS}, we then
adapt to the case of correlated potentials
the elementary and short argument of \cite{JS} showing that
positivity of the Lyapunov exponents implies at most logarithmic growth of
quantum dynamics and hence, by Guarneri's inequality \cite{Gua}, 
zero Hausdorff dimension of the spectral measures. Even though this is a
weaker localization result than pure-point spectrum with exponential localized
eigenfunctions, it proves the behavior which is stable under perturbation and 
we hence consider, as argued in \cite{JS}, that it already
captures the physically relevant effect. In the next section, the
results and the precise hypothesis are described and discussed 
in detail. The other sections
contain the proofs.

\vspace{.2cm}

\noindent {\bf Acknowledgment:} This work was supported by the DFG.

\section{Set-up and results}
\label{sec-results}

In order to fix terms and notations, we have to begin by reviewing some basic
definitions of symbolic dynamics and strong mixing \cite{Bow,PP}.
Let $\Sigma$ be a countable set furnished with the discrete topology. We
designate a reference element $0\in\Sigma$. For any subset
$I\subset \ZZ$ and $\omega=(\sigma_n)_{n\in\ZZ}\in\Omega$, let us define
$$
\pi_I(\omega)
\;=\;
(\hat\sigma_n)_{n\in\ZZ}
\;,
\qquad
\hat\sigma_n
\,=\,0
\quad\text{for} \quad n \not\in I\;, 
\qquad
\hat\sigma_n 
\,=\, 
\sigma_n
\quad\text{for}\quad n \in I
\;.
$$
For a bounded, measurable function $g:\Omega\to V$ into 
a real, normed vector space $(\VM,\|\,.\,\|)$, the variation on $I$ 
is defined by
$$
\Var_I(g)
\;=\;
\sup_{\pi_I(\omega)=\pi_I(\omega')}\,
\|g(\omega)\,-\,g(\omega')\|\;.
$$
Then $g$ is called
quasi-local with rate $0 < r < 1$ if and only if 
there exists a constant $C=C(g)$ 
such that, for any $m,n \geq 1$,
\begin{equation}
\label{eq-quasilocal}
\Var_{[-m,n]}(g)
\;\leq\;
C(g)\,r^{m\wedge n}
\,,\qquad
m\,\wedge\,n
\;=\; \min\{m,n\}
\end{equation}
The set of all quasi-local functions with rate $r$ 
is denoted by $\Qq_r(\VM)$. 
%
%

\vspace{.2cm}

Next let us state precisely the strong mixing hypothesis used in this work. 
For $m<n$ and $a_k\in\Sigma$ with $m\leq k
\leq n$, the associated cylinder set is denoted by
$A_{m,n}=A_{m,n}(a_m,\ldots,a_n)=
\{\omega=(\sigma_k)_{k\in\ZZ}\,|\, \sigma_k=a_k,\, m\leq k \leq n\}$.
Then the invariant measure $\PP$ on the shift space $(\Omega,\ZZ)$
is said to satisfy a power law $\psi$-mixing \cite{Bra} with
exponent $\alpha>0$ if there is a constant $C>0$ 
such that for all $k<l<m<n$ and all $A_{k,l},\, A_{m,n}$, one has
\begin{equation}
\label{eq-cylinder}
\Big|\,
\PP(A_{k,l}\cap\,A_{m,n})
\;-\;\PP(A_{k,l})\PP(A_{m,n})\,\Big|
\;\leq\;
C\,\PP(A_{k,l})\PP(A_{m,n})\;|m-l|^{-\alpha}\;.
\end{equation}
Equivalently, for any $\pi_{[k,l]}$-measurable
function $g_1$ and $\pi_{[m,n]}$-measurable function $g_2$ holds 
\begin{equation}
\label{eq-expmixing}
\Big|\,
\EE(g_1\, g_2)\,-\,\EE(g_1)\,\EE(g_2) \,\Big|
\;\leq\;
C\,\EE(|g_1|)\,\EE(|g_2|)\,|m-l|^{-\alpha}\;,
\end{equation}
where $k<l<m<n$ and $C$ as above. This also implies ergodicity. 
Examples when \eqref{eq-expmixing} holds are given in Remark 1 and 2 below,
after the main results are stated. Averages over $\omega$ 
w.r.t. $\PP$ are denoted by $\EE$, or also by $\EE_\omega$ if the dependence on
$\omega$ is retained in the integrand. Furthermore, the set of
centered quasi-local functions will be denoted by
$\Qq^0_r(\VM)=\{g\in\Qq_r(\VM)\,|\, \EE(g)=0\} $.

\vspace{.2cm}

Throughout we suppose that the potential in \eqref{eq-Jacobi} is 
given by a centered real-valued 
quasi-local function $V \in \Qq^0_r(\RR)$. It is
well-known and verified in Lemma~\ref{lemma-weakcorrel} that 
\eqref{eq-expmixing} implies the decay of correlations
$|\EE(V(\omega) V(S^n\omega))|\leq C |n|^{-\alpha}$ 
for some constant $C$. For $\alpha>1$, one can hence define
its (positive) spectral density $D_V(k)$ at $k\in[0,2\pi)$:
$$
D_V(k)
\;=\;
\sum_{n\in\ZZ}\,e^{\imath kn}\,\EE_\omega\bigl(V(\omega)V(S^n\omega)\bigr)
\;=\;
\lim_{N\to\infty}\,\frac{1}{N}\,
\EE_\omega\left(\Bigl|\sum_{n=0}^{N-1}e^{\imath kn} V(S^n\omega)\Bigr|^2\right)
\;.
$$

As final preparation let us recall the definition of the Lyapunov exponent
$\gamma_\lambda(E)$ at energy $E\in\CC$ associated to \eqref{eq-Jacobi}. If
the transfer matrices are defined by
\begin{equation}
\label{eq-transfer}
\Tt^E_{\lambda,\omega}
\;=\;
\left( \begin{matrix}
E-\lambda V(\omega) & -1 \\ 1 & 0
\end{matrix} \right)
\;\in\;\Qq_r\bigl({\rm SL}(2,\RR)\bigr)
\;,
\end{equation}
then
$$
\gamma_\lambda(E)
\;=\;
\lim_{N\to\infty}\;
\frac{1}{N}\;
\EE_\omega\;\log\left(\Bigl\|\prod^N_{n=1}\Tt^E_{\lambda,S^n\omega}
\Bigr\|\right)
\;.
$$

The main result of Chulaevski and Spencer \cite{CS} is that
for $\alpha>2$ and at an energy $E=2 \cos(k)$ 
in the spectrum $[-2,2]$ of the discrete Laplacian, but
away from the band edges $E=-2,2$ and the band center $E=0$ such that the
distance $d(k)$ of $k$ from $0\!\!\mod \!\frac{\pi}{2}$ is positive:
\begin{equation}
\label{eq-Thouless}
\gamma_\lambda(E)
\;=\;\lambda^2\;
\frac{D_V(k)}{8\,\sin^2(k)}
\;+\;
\Oo\Bigl(\frac{\lambda^{\frac{3\alpha+2}{\alpha+2} }}{d(k)}\Bigr)
\;.
\end{equation}
As we need to build up the whole formalism anyway, the main element of the 
proof of
\eqref{eq-Thouless} is reproduced in Section~\ref{sec-CS}. 
As indicated, the control of the error terms breaks down at the band edges  
and the band
center. Our first result provides  perturbative formulas for the Lyapunov
exponent at these
energies, generalizing respectively our prior results for independent
potential values \cite{S,SS}.

\begin{theo}
\label{theo-result}
Assume $\alpha>2$, $D_V(0)>0$ and $D_V(\pi)>0$ {\rm(}the latter is only needed
for {\rm(i))}. 

\vspace{.1cm}

\noindent {\rm (i)} 
The Lyapunov exponent near the band center $E=0$ is given by
\begin{equation}
\gamma_\lambda(\epsilon\lambda^2)
\;=\;\lambda^2\;
\frac{D_V(\pi)}{8}\,
\int_0^\pi {\rm d}\theta\,\rho_\epsilon(\theta)\,(1+\cos(4\theta))
\;+\;\Oo(\lambda^{\frac{3\alpha+2}{\alpha+2}})\;,
\label{eq-resultbandcenter}
\end{equation}
where $\rho_\epsilon$ is a $\pi$-periodic smooth probability density. 

\vspace{.1cm}

\noindent {\rm (ii)} 
Up to errors of order $\Oo(\lambda^{\frac{3\alpha+2}{3\alpha+6}})$,
the Lyapunov exponent near the upper band edge $E=2$ is given by
\begin{equation}
\gamma_\lambda(2+\epsilon\lambda^{\frac{4}{3}})
\,=\,
\lambda^{\frac{2}{3}}\left(
\frac{1-\epsilon}{2}\int {\rm d}\theta\,\rho_\epsilon(\theta)\,\sin(2\theta)+
\frac{D_V(0)}{8}\int{\rm d}\theta \rho_\epsilon(\theta)\,
\bigr(1+2\cos(2\theta)+\cos(4\theta)\bigr)\right),
\label{eq-resultbandedge}
\end{equation}
where $\rho_\epsilon$ is a $\pi$-periodic smooth probability density
written out explicitly in {\rm Section~\ref{sec-application}}.
The same formula holds at the lower band edge $E=-2$. 
\end{theo}

\vspace{0.1cm}

The formulas \eqref{eq-Thouless}, \eqref{eq-resultbandcenter} 
and \eqref{eq-resultbandedge} combined allow to study the Lyapunov exponents
at all energies $[-2,2]$. In order to assure positivity for $\lambda>0$, one
first has to check that the spectral density is positive ({\it cf.}
Remark~\ref{rem-density} 
below) and then prove that the integrals appearing in
\eqref{eq-resultbandcenter}  and \eqref{eq-resultbandedge} are positive. This
is immediate for \eqref{eq-resultbandcenter}. For \eqref{eq-resultbandedge}
we could not produce an analytic proof, but, given the explicit formula 
\eqref{eq-rhoformula} for $\rho_\epsilon$, one can readily do a numerical
evaluation. 

\vspace{0.1cm}

Nevertheless, the three formulas are not yet sufficient to prove uniform 
positivity of
the Lyapunov exponent on the whole spectrum for some fixed small, but positive
value of $\lambda$. Indeed, for once the non-random spectrum
$\sigma(H_{\lambda,\omega})$ may (and typically will) fill the whole interval
$[-2-\lambda \|V\|_\infty,2+\lambda \|V\|_\infty]$ where 
$\|V\|_\infty=\PP\!-\!\esssup |V(\omega)|$ (use approximate eigenfunctions as
Weyl sequences in order to show this). For an energy $2+\epsilon\lambda$,
$\epsilon>0$, the asymptotics 
\eqref{eq-resultbandedge} then says nothing. However,
one can combine the techniques of this work with those of \cite{SS} in order to
prove, as in the case of independent potential values 
\cite[Section~8]{SS},
\begin{equation}
\label{eq-bandedge-hyp}
\gamma_\lambda(2+\epsilon\lambda^\eta)
\;=\;
\sqrt{\epsilon \lambda^\eta}\; 
+\;\Oo(\lambda^{1-\frac{\eta}{4}},
\lambda^{\frac{7\eta}{4}-1},
\lambda^{\frac{\eta}{4}\frac{3\alpha+2}{\alpha+2}})\;,
\qquad
\epsilon>0\;,
\end{equation}
where $\frac{4}{5}< \eta<\frac{4}{3}$ is such that the error terms are of
lower order than $\lambda^{\frac{\eta}{2}}$ (in particular, $\eta=1$ is
allowed for $\alpha$ sufficiently large).
Moreover, the
formulas  \eqref{eq-Thouless} and \eqref{eq-resultbandedge} do not imply
positivity of the Lyapunov exponent at a fixed $\lambda$ for all energies
in $[2-\lambda,2)$ because the error term in \eqref{eq-Thouless} explodes as
one approaches the band edge. However, once again one can transpose 
\cite[Section~8]{SS}
to the case of a strongly mixing potential:
\begin{equation}
\label{eq-bandedge-ellipt}
\gamma_\lambda(2-\epsilon\lambda^\eta)
\;=\;
\lambda^{2-\eta}\;\frac{D_V(0)}{8\,\epsilon} 
+\;\Oo(\lambda^{4-\frac{5\eta}{2}},
\lambda^{\frac{3\eta}{2}}, 
\lambda^{(1-\frac{\eta}{2})\frac{3\alpha+2}{\alpha+2}})\;,
\qquad
\epsilon>0\;,
\end{equation}
where again $\frac{4}{5}< \eta<\frac{4}{3}$ has to assure that the error terms
are subdominant. 
A careful analysis now allows to show (modulo the issues discussed above) 
that for $\lambda$ sufficiently small the Lyapunov exponent is positive on 
$[2-c,2+c]$ for $c>0$. 
We do not provide the detailed argument here, but do claim to have presented  
all the essential ingredients in order 
to complete it. Similarly, by analyzing the
Lyapunov exponent $\gamma_\lambda(\epsilon\lambda^\eta)$, $1\leq\eta\leq 2$,
using the techniques of \cite[Section 5.1]{S} or \cite[Section 5]{SS}, one can
show that the Lyapunov exponent is positive near the band center for $\lambda$
sufficiently small.

\vspace{.2cm}

Let us now assume that uniform positivity of the Lyapunov exponent has been
verified for all energies in the spectrum, either by the above or some other
argument, and then deduce localization estimates from this. One standard way to
quantify the spreading (delocalization) of an
initially localized wave packet $|0\rangle$ under the quantum mechanical 
time evolution 
$e^{-\imath tH_{\lambda,\omega}}$ is to consider the growth of 
(time and disorder averaged) moments of 
the position operator $X$ on $\ell^2(\ZZ)$: 
\begin{equation}
\label{eq-moment}
M^q_T
\;=\;
\int^\infty_{0}\frac{dt}{T}\;e^{-\frac{t}{T}}\;\EE_\omega\;
\langle 0 |
\,e^{\imath H_{\lambda,\omega} t}\,|X|^q\, e^{-\imath H_{\lambda,\omega} t}\,
|0\rangle
\mbox{ , }
\qquad
q>0
\mbox{ . }
\end{equation}
Boundedness of $M^q_T$ uniformly in time is called dynamical
localization. Logarithmic growth in time as obtained in the following theorem 
is quite close to that.

\begin{theo}
\label{theo-logbound}
Consider an ergodic family of Jacobi matrices
$(H_{\lambda,\omega})_{\omega\in\Omega}$ of the form  {\rm \eqref{eq-Jacobi}}
with a quasi-local potential $V$ and an invariant measure $\PP$ satisfying 
{\rm \eqref{eq-expmixing}} with $\alpha>0$. Suppose that the
spectrum is included in an open interval $(E_0,E_1)$ on which the 
Lyapunov exponent is uniformly positive:
\begin{equation}
\label{eq-lyaplower}
\gamma_\lambda(E)
\;\geq\;
\gamma_0\;>\;0\;,
\qquad
E\in (E_0,E_1)\;.
\end{equation}
Then for any $\beta>2$ there exists a constant $C(\beta,q)$ such that
\begin{equation}
\label{eq-logupper}
M_T^q
\;\leq\;
(\log T)^{q\beta}\;+\;C(\beta,q)
\mbox{ . }
\end{equation}
Furthermore, the Hausdorff dimension of the 
spectral measure of $H_{\lambda,\omega}$ vanishes $\PP$-almost
surely.
\end{theo}

The elementary proof (fitting on 4-5 pages) of \eqref{eq-logupper}
is almost completely contained in \cite{JS}.
It is therefore not reproduced here, but we discuss in detail in
Section~\ref{sec-bound} the only step that has to be modified.
As already indicated in the introduction, the last statement then follows
directly from Guarneri's inequality \cite{Gua}.

\vspace{.2cm}

Now follow remarks on when the hypothesis of the above theorems are
satisfied. 

\begin{rem} {\rm
The strong mixing condition \eqref{eq-expmixing} clearly holds if  $\PP$ is
the product measure of some probability 
measure on $\Sigma$, because the functions $g_1$
and $g_2$ are then independent. The mixing condition also holds if $\PP$ stems
from a Markov process given by a stochastic kernel having only
one invariant measure on a countable set $\Sigma$. 
Then the decay on the r.h.s. of
\eqref{eq-expmixing} is actually exponential, with rate given by the
Perron-Frobenius gap of the stochastic kernel. Yet more general, let us
consider a hyperbolic dynamical system $(X,T)$ (Axiom A) given by a map
$T:X\to X$. Then one has a finite Markov
partition $\Sigma$, with associated symbolic dynamics $(\Omega,S)$ \cite{Bow},
and there
is a wealth of so-called Gibbs measures associated to H\"older continuous
({\it i.e.} quasi-local) functions which all satisfy \eqref{eq-expmixing}
with an exponential mixing rate \cite[Proposition 2.4]{Bow}.
Two standard examples of this type already cited in \cite{CS} 
are the period doubling map and the Arnold cat maps.   
Moreover, if the phase space $X$ is a
manifold, then any differentiable real function on this manifold gives rise to
a quasi-local potential under the coding map. For all these examples with
exponential $\psi$-mixing, the error bounds in \eqref{eq-Thouless},
\eqref{eq-resultbandcenter} and \eqref{eq-resultbandedge} are given by the
error bounds of the independent case \cite{PF,S,SS} multiplied by
$\log^2(\lambda)$. The error bounds in the independent case are recovered by
sending $\alpha\to\infty$ in \eqref{eq-Thouless},
\eqref{eq-resultbandcenter}, \eqref{eq-resultbandedge},
\eqref{eq-bandedge-hyp} and \eqref{eq-bandedge-ellipt}
}
\end{rem}

\begin{rem} {\rm
Concrete examples of dynamical systems $(X,T)$ 
having not an exponential, but only a 
power law decay in \eqref{eq-expmixing} have only be analyzed more recently.
Necessarily $T$ is then not uniformly
hyperbolic, but it is supposed to have only a few parabolic points. 
Such examples can be constructed even if $X$ is an interval, but the
invariant measure then has a non-normalizable density w.r.t. the
Lebesgue measure.
It is, however, possible to construct a symbolic dynamics over a countable
alphabet $\Sigma$ which then has an shift-invariant propability measure $\PP$
satisfying the strong mixing estimates \eqref{eq-cylinder} and
\eqref{eq-expmixing}.
Instead of producing a long citation list, we refer to the references in
\cite{Gou} which contains a proof of \eqref{eq-expmixing} for several
concrete examples. It is precisely in order to deal with these cases at the
verge that we bothered to work with \eqref{eq-expmixing} instead of
exponential mixing.
}
\end{rem}

\begin{rem} 
\label{rem-density} 
{\rm
The positivity of the spectral density $D_V(k)$ can for some examples be
checked by an explicit calculation, but there are also further techniques
available in order to verify this \cite{Bra}. 
The case of $D_V(0)$ is particularly well studied
because of its importance for central limit theorems \cite{PP,Gou}. For
the Gibbs measures of Remark~1 and the examples of Remark~2,
$D_V(0)=0$ holds if and only if $V=v\circ S
-v$ is a cocycle given by another quasi-local function $v$. By suspension, one
can deal similarly with $k=\pi$ and actually any rational $\frac{k}{2\pi}$.  
}
\end{rem}

\begin{rem} {\rm
The above results transpose if $\ZZ$ is replaced by $\NN$, namely for 
$\Omega=\Sigma^\NN$ furnished with the left shift and $H_{\lambda,\omega}$
acts on $\ell^2(\NN)$.
As the inverse $S^{-1}$ of the left shift operator is not defined
in that case, one needs to replace in all proofs functions like $g\circ S^{-n}$
for $n>0$ by ${(U^*)}^n g $, where $U^*$ is the 
$L^2(\Sigma^\NN,\PP)$-adjoint operator
of $U: g \mapsto g\circ S$. 
}
\end{rem}

\section{Anomalies at band center and band edge}
\label{sec-normalform}

Let us begin by recalling that the transfer matrix
$\Tt^E_{\lambda,\omega}\in\, $SL$(2,\RR)$  given in \eqref{eq-transfer} is
elliptic for an
energy $E=2\cos(k)\in (-2,2)$ and $\lambda=0$, 
and it can hence, to zeroth order in $\lambda$,  
be transformed into a rotation. 
More explicitly,
\begin{equation}
\label{eq-expan0}
M\Tt^E_{\lambda,\omega}M^{-1}\;=\;
R_k\,\left(\one + \lambda\;\frac{V(\omega)}{\sin(k)}\,
\left( \begin{matrix}
0 & 0 \\ 1 & 0
\end{matrix} \right)
\;\right)
\;,
\end{equation}
where
$$
R_k
\;=\;
\left( \begin{matrix}
\cos(k) & -\sin(k) \\ \sin(k) & \cos(k)
\end{matrix} \right)
\;,
\qquad
M
\;=\;
\frac{1}{\sqrt{\sin(k)}}\;
\left( \begin{matrix}
\sin(k) & 0 \\ -\cos(k) & 1
\end{matrix} \right)
\;.
$$
In the next section we will consider the action of the matrix
\eqref{eq-expan0} on the real
projective line, which is identified with a circle. To lowest order
$\lambda^0$, this action induced by \eqref{eq-expan0} is then a rotation on the
circle.  For irrational $\frac{k}{2\pi}$, 
there is a unique
invariant measure given by the Lebesgue measure. For rational 
$\frac{k}{2\pi}=\frac{p}{q}$ at least Birkhoff sums of harmonics of order
lower than $q$ vanish. 

\vspace{.2cm}

At the band center $k=\frac{\pi}{2}$, the square of the 
transfer matrix \eqref{eq-expan0} (note that $M=\one$ here) 
is the unit matrix and one can
only control the lowest order harmonic, which turns out not to be sufficient
for the calculation of the Lyapunov exponent. It is then more convenient to
consider directly the square of the transfer matrix
\begin{eqnarray}
\Tt^{\epsilon\lambda^2}_{\lambda,S\omega}\,
\Tt^{\epsilon\lambda^2}_{\lambda,\omega}
& = & 
- \left( \begin{matrix}
1-\lambda^2 V(\omega) V(S\omega) & \epsilon \lambda^2-\lambda V(S\omega) \\
-\epsilon \lambda^2 + \lambda V(\omega)& 1 
\end{matrix}\right)
\,+\, \Oo(\lambda^{3}) \nonumber
\\
& &
\label{eq-normform-bandcenter}
\\
&=& 
-\exp\left(
\lambda\left(\begin{matrix}
0 & -V({S^n\omega}) \\
V(\omega) & 0 \end{matrix} \right) 
\,+\,\frac{\lambda^2}{2} \left( \begin{matrix}
-V(\omega) V({S\omega}) & 2\epsilon \\
-2\epsilon & V_\omega V({S\omega}) \end{matrix} \right)
\,+\,\Oo(\lambda^{3})\,  \right)\;,\nonumber
\end{eqnarray}
and to group
the coordinates of $\omega$ in pairs and consider
$\tilde{\Omega}=\tilde{\Sigma}^\ZZ$ where
$\tilde{\Sigma} = \Sigma \times \Sigma$, and furnish it with a probability
$\tilde{\PP}$ naturally induced by $\PP$. Again the suspension
$(\tilde{\Omega},\tilde{\PP})$
is a shift space with power law mixing. However, the matrix 
\eqref{eq-normform-bandcenter} is now in the form of an anomaly as discussed
at the end of this section.

\vspace{.2cm}

At a band edge, {\it e.g.} $E=-2$ and $k=\pi$, the basis change in
\eqref{eq-expan0} becomes singular and one has a Krein collision. Nevertheless,
the transfer matrix at $\lambda=0$ can be transformed into a non
diagonalizable Jordan normal form:
$$
N\,\Tt^{-2+\epsilon\lambda^{\frac{4}{3}}}_{\lambda,\omega}\,N^{-1}
\;=\;
- \left( \begin{matrix}
1+\lambda V(\omega) - \epsilon \lambda^{\frac{4}{3}} & 1 \\
\lambda V(\omega) - \epsilon \lambda^{\frac{4}{3}} & 1
\end{matrix} \right)\;,\quad
N\;=\; \left(\begin{matrix}
1 & 0 \\ -1 & 1
\end{matrix} \right)\;.
$$
Let us further conjugate this
matrix by  $N_{\lambda}=\left( \begin{matrix} 
\lambda^{\frac{2}{3}} & 0 \\ 0 & 1 \end{matrix}\right)$ in order to get again
an anomaly ({\it cf.} \cite{SS} for a motivation of this conjugation): 
\begin{equation}
\label{eq-bandedgeanomaly}
N_{\lambda}\,N\,
\Tt^{-2+\epsilon\lambda^{\frac{4}{3}}}_{\lambda,\omega}\,
N^{-1}\,N^{-1}_{\lambda}
\;=\;
-\,\exp\left(\,
\lambda^{\frac{1}{3}}\,\left( \begin{matrix}
0 & 0 \\ V(\omega) & 0 \end{matrix} \right)\,+\,
\lambda^{\frac{2}{3}}\, \left(\begin{matrix}
0 & 1 \\ -\epsilon & 0 \end{matrix} \right)\,+\,
\Oo(\lambda)\,
\right)\;.
\end{equation}

Resuming, after adequate basis change and possibly regrouping of terms,
one has to study in each of the three
situations \eqref{eq-expan0}, 
\eqref{eq-normform-bandcenter} and \eqref{eq-bandedgeanomaly}
families of random matrices
$(T_{\lambda,\omega})_{\lambda\geq 0, \omega\in\Omega}\,
\in\,\Qq_r\bigl({\rm SL}(2,\RR)\bigr)$ of the following form:
\begin{equation}
\label{eq-expan}
T_{\lambda,\omega}
\;=\;
\pm\,R_k\;
\exp\left(
\lambda^{\eta}\,P_{1,\omega}\;+\;
\lambda^{2\eta}\,P_{2,\omega}\;+\;\Oo(\lambda^{3\eta})\;
\right)
\;
\end{equation}
where $\eta>0$, 
$P_{j,\omega}\in\Qq_r\bigr({\rm sl}(2,\RR)\bigr)$ for $j=1,2$, 
$\EE(P_{1,\omega})=0$ and
the error term $\Oo(\lambda^{3\eta})$ is uniformly bounded
({\it i.e.} the bound is $\omega$-independent).
If $k=0,\pi$, namely at a band center \eqref{eq-normform-bandcenter} and a
band edge \eqref{eq-bandedgeanomaly}, such a family is said to have an
anomaly of second order \cite{S,SS}. In the following sections, we
treat general families of the form \eqref{eq-expan}, and then go back to the
explicit cases in Section~\ref{sec-application} in order to complete the proof
of Theorem~\ref{theo-result}.

\section{Phase shift dynamics}
\label{sec-dynamics}

The bijective action $\Ss_T$ of a matrix $T\in\mbox{SL}(2,\RR)$ on 
$S^1_\pi= \RR/\pi\ZZ=[0,\pi)$ is given by

\begin{equation}
\label{eq-action}
e_{\Ss_T(\theta)}
\;=\;
\pm\;\frac{Te_{\theta}}{
\|Te_{\theta}\|}
\;,
\qquad
\;\;\;\;\;
e_\theta
\;=\;
\left(
 \begin{array}{cc} \cos(\theta) \\ \sin(\theta)
\end{array}
\right)
\mbox{ , }
\;\;
\theta\in S^1_\pi
\mbox{ , }
\end{equation}

\noindent with an adequate choice of the sign. 
This defines a group action, namely
$\Ss_{TT'}=\Ss_{T}\Ss_{T'}$.
In order to shorten notations, we write 
$\Ss_{\lambda,\omega}=\Ss_{T_{\lambda,\omega}}$ where $T_{\lambda,\omega}$ is
of the form \eqref{eq-expan}. One thus has
$\Ss_{\lambda,\omega}(\theta)=\theta+k+\Oo(\lambda)$. 

\vspace{.2cm}

Given an initial angle $\theta_0$ and iterating this dynamics by
the left shift on $\Omega$ defines
a stochastic process $\theta_n(\omega)$, also simply denoted by $\theta_n$
below:
\begin{equation}
\label{eq-randomdyn}
\theta_0(\omega)
\;=\;
\theta_0
\,,
\qquad
\theta_{n+1}({\omega})
\;=\;
\Ss_{{\lambda,S^n\omega}}(\theta_n(\omega))
\;.
\end{equation}

In order to analyze the dynamics in more detail, let us introduce for $j=1,2$
the trigonometric polynomials
\begin{equation}
\label{eq-polynom}
p_{j,\omega}(\theta)
\;=\;
\Im m \left(\frac{\langle v|P_{j,\omega}|e_\theta\rangle}{\langle
v|e_\theta\rangle}
\right)
\;,
\qquad 
v\;=\;
\frac{1}{\sqrt{2}}\left(\begin{array}{c}
1\\-\imath\end{array}\right)
\;.
\end{equation}
One then has
({\it cf.} \cite{SS} for details)
\begin{equation}
\label{eq-polydef}
\Ss_{\lambda,\omega}(\theta)
\;=\;
\theta\,+\,k\,+\,\sum_{j=1}^2 \lambda^{j\eta}\,p_{j,\omega}(\theta)
\,+\,\frac{1}{2}\,\lambda^{2\eta}\,
p_{1,\omega}\,
\partial_\theta\, p_{1,\omega}(\theta)
\,+\,\Oo(\lambda^{3\eta})
\;.
\end{equation}

Due to Lemma~3 of \cite{JSS} and a telescoping argument, the 
Lyapunov exponent $\gamma(\lambda)$ characterizing the exponential growth
of the products of matrices in the 
ergodic family  $(T_{\lambda,S^n\omega})_{n\geq 0}$ is given by
\begin{equation}
\label{eq-lyap}
\gamma(\lambda) 
\;=\; 
\lim_{N \to\infty}\,\frac{1}{N}
\,\EE_{\theta_0}\EE_\omega\,\sum_{n=0}^{N-1}
\,\log(\|T_{\lambda,S^n\omega} e_{\theta_n(\omega)}\|) \;,
\end{equation}
where $\EE_{\theta_0}$ denotes an average over the initial condition
$\theta_0$ w.r.t. an arbitrary continuous probability measure on $S^1_\pi$. 
As our interest is perturbation theory for $\gamma(\lambda)$
w.r.t. $\lambda$, we
shall need the following expansions for the summands of \eqref{eq-lyap}
({\it e.g.} \cite{SS} contains the algebraic proof):

\begin{lemma}
\label{lem-coefficients}
Set
$$
\alpha_{j,\omega}
\;=\;
\langle v|\,P_{j,\omega}\,|v\rangle
\;,
\qquad
\beta_{j,\omega}
\;=\;
\langle \overline{v}|\,P_{j,\omega}\,|v\rangle
\;,
\qquad
\gamma_{j,\omega}
\;=\;
\langle \overline{v}|\,|P_{j,\omega}|^2\,|v\rangle
\;
$$
Then $p_{j,\omega}(\theta)=\Im m(
\alpha_{j,\omega}-\beta_{j,\omega}e^{2\imath\theta})$. Furthermore,
\begin{equation}
\label{eq-expan1}
\log(\|T_{\lambda,\omega} e_{\theta}\|)
\;=\;
\Re e\,\Bigl(\sum_{j= 1}^2
\lambda^{j\eta} \,
\beta_{j,\omega}\,e^{2\imath\theta}
\,+\,
\frac{\lambda^{2\eta }}{2}\,
\bigl(
|\beta_{1,\omega}|^2
\,+\,
\gamma_{1,\omega}\,e^{2\imath\theta}
\,-\,
\beta_{1,\omega}^2\,e^{4\imath\theta}
\bigr)
\Bigr)\,+\,\Oo(\lambda^{3\eta})\;.
\end{equation}
\end{lemma}

\vspace{.2cm}

Formula \eqref{eq-lyap} and also its perturbative evaluation based on
\eqref{eq-expan1} hence leads us to consider
sums of the type
\begin{equation}
\label{eq-quasiosci}
\hat{I}_N(\Gg)
\;=\;
\frac{1}{N}\,\EE_\omega\,\sum_{n=0}^{N-1} \Gg(S^n\omega,\theta_n(\omega))
\;,
\qquad
\hat{I}(\Gg)\;=\;
\lim_{N\to\infty}\,\hat{I}_N(\Gg)
\;,
\end{equation}
\noindent for functions $\Gg$ on $\Omega\times S^1_\pi$ of the type
$\Gg(\omega,\theta)=\sum_j g_j(\omega)f(\theta)$.
More explicitly, the above lemma shows that one only
needs functions of the form
$g(\omega) e^{2\imath\theta}$ and $g(\omega) e^{4\imath\theta}$ with
$g \in \Qq_r(\CC)$. For a $\pi$-periodic function $f \in C(S^1_\pi)$, we also
introduce 
$$
I_N(f)
\;=\;
\frac{1}{N}\,\EE\,\sum_{n=0}^{N-1} f(\theta_n)
\;,
\qquad
I(f)
\;=\;
\lim_{N\to\infty}\,I_N(f)\;,
$$
This is a Birkhoff sum of the process $\theta_n=\theta_n(\omega)$.
In the sum \eqref{eq-quasiosci} there is, moreover, an 
explicit dependence of $\Gg$ on $\omega$, hence let us use the term
Birkhoff-like sums for the 
sums $\hat{I}(\Gg)$. 

\section{From Birkhoff-like sums to Birkhoff sums}
\label{sec-birklike-birk}

The aim of this section is, as indicated in the title, to reduce the
perturbative  evaluation of the Birkhoff-like sums \eqref{eq-quasiosci} 
to the evaluation of
Birkhoff sums by invoking the correlation decay \eqref{eq-expmixing}. 

\begin{proposi}
\label{prop-birklike-birk}
Suppose $\alpha>2$.
Let $g \in \Qq_r(\CC)$ and $f\in C^2(S^1_\pi)$. 
Define $\Gg(\omega,\theta) = g(\omega) f(\theta)$.
Then
\begin{equation}
\label{eq-birklike-birk}
\hat{I}(\Gg)
\;=\;
\EE(g)\,I(f)\,+\,\Oo(\lambda^{\eta\frac{\alpha}{1+\alpha}})
\;.
\end{equation}
If $\EE(g)=0$, one has the following convergent expression for the next higher
order contribution:
\begin{equation}
\label{eq-birklike-birk1}
\hat{I}(\Gg)
\;=\;\,
\lambda^{\eta}\sum_{j=1}^\infty I(f_{j})
\,+\, \Oo(\lambda^{\eta\frac{2\alpha}{2+\alpha}})\;, 
\qquad
f_{j}(\theta)
\;=\;
\EE_\omega
\bigl(\,g(S^j \omega) p_{1,\omega}(\theta)\bigr)\,f'(\theta)
\;.
\end{equation}
\end{proposi}

The first lemma needed for the proof is mainly contained in \cite{CS}.
We provide a few more details of the proof and use the notations of this work.

\begin{lemma}
\label{lemma-change-theta}
One has for $m,n\geq 1$ 
$$
\Var_{[-m-n,n+m]}(\theta_n)
\;\leq\;
\Oo(r^m\lambda^{\eta})
\;.
$$
\end{lemma}

\noindent {\bf Proof.} Using equation \eqref{eq-randomdyn},
$$
\bigr|\theta_{n+1}(\omega)\,-\,\theta_{n+1}(\omega')\bigl|
\;\leq\;
\bigl|\Ss_{\lambda,S^n\omega}(\theta_n(\omega))-
\Ss_{\lambda,S^n\omega}(\theta_n(\omega'))\bigr|\,+\,
\bigl|\Ss_{\lambda,S^n\omega}(\theta_n(\omega'))-
\Ss_{\lambda,\omega'}(\theta_n(\omega'))\bigr|
\;,
$$
one deduces
\begin{equation}
\label{eq-vartheta1}
\bigr|\theta_{n+1}(\omega)\,-\,\theta_{n+1}(\omega')\bigl|
\;\leq\;
\left(\sup_{\omega,\theta} \;|\Ss'_{\lambda,\omega}(\theta)|\right)\;
|\theta_n(\omega)-\theta_n(\omega')|\,+\,
\sup_{\theta} \bigl|\Ss_{\lambda,S^n\omega}(\theta)-
\Ss_{\lambda,S^n\omega'}(\theta)\bigr|\;.
\end{equation}
Using the estimate
$$
\left\| \frac{x}{\|x\|}-\frac{ x'}{\| x'\|}\right\|
\;=\;
\left\|\frac{x- x'}{\|x\|}\,+\,{x'}\left(
\frac{\|{x'}\|-\|x\|}{\|x\|\,\|{x'}\|}\right)
\right\|
\;\leq\; 
\frac{2}{\|x\|}\,\|x\,-\,{x'}\|
$$
and the definition of $\Ss_{\lambda,\omega}$, it follows
$$
\left\|e_{\Ss_{\lambda,\omega}(\theta)}-e_{\Ss_{\lambda,\omega'}(\theta)}
\right\|
\;\leq\;
\frac{2}{\|T_{\lambda,\omega}e_\theta\|}
\,\left\|T_{\lambda,\omega}-T_{\lambda,\omega'} \right\|
\;\leq\;
2 \left(\sup_{\omega} \|T^{-1}_{\lambda,\omega}\|\right)
\,\left\|T_{\lambda,\omega}-T_{\lambda,\omega'} \right\|\;.
$$
This implies
\begin{equation}
\label{eq-vartheta2}
\sup_{\theta}\; \bigl|\Ss_{\lambda,S^n\omega}(\theta)-
\Ss_{\lambda,S^n\omega'}(\theta)\bigr|
\;\leq\;
C_1\,\lambda^\eta\,\| Q_{\lambda,\omega}-Q_{\lambda,\omega'} \|
\end{equation}
where $C_1$ is a constant and
$T_{\lambda,\omega}={\bf 1}+\lambda^\eta Q_{\lambda,\omega}$
for a matrix-valued function $Q_{\lambda,\omega}$ that is
analytic in $\lambda^\eta$ and uniformly quasi-local for small $\lambda$
({\it i.e.} the constant and rate is $\lambda$-independent).
Furthermore, one has
$$
\sup_{\omega,\theta} |\Ss'_{\lambda,\omega}(\theta)| 
\;\leq\;
1\,+\,C_2\,\lambda^\eta
$$
for $\lambda$ sufficiently small and some constant $C_2$.
Applying this and \eqref{eq-vartheta2} to \eqref{eq-vartheta1}
one gets
$$
\Var_I(\theta_{n+1})
\;\leq\;
(1+C_2\,\lambda^\eta)\;\Var_I(\theta_n)
\;+\;
C_1\,\lambda^{\eta} \, \Var_I(Q_{\lambda,S^n\omega})\;.
$$
Iterating this estimate and using $\Var_I(\theta_0) = 0$, it follows that
\begin{eqnarray*}
\Var_{[-m-n,n+m]}({\theta_n}) 
& \leq &
\sum_{j=1}^n (1+C_2\lambda^{\eta})^{j-1}\,C_1\lambda^\eta
\Var_{[-m-n,n+m]}(Q_{\lambda,S^{n-j}\omega})
\\
& \leq &
C_1\lambda^{\eta}\,C_3\,r^{m+1}\,\sum_{j=0}^\infty(1+C_2\lambda^{\eta})^j r^j
\;=\;
\Oo(\lambda^{\eta} r^m)
\end{eqnarray*}
for $\lambda$ sufficiently small.
\hfill $\Box$

\vspace{.2cm}

In order to state 
the next two lemmata, we introduce the following notation extending
\eqref{eq-quasiosci}: 
$$
\hat{I}^m_N(\Gg)
\;=\;
\frac{1}{N}\,\EE\,\sum_{n=0}^{N-1} \Gg(S^{m+n}\omega,\theta_n(\omega))
\;,
\qquad
\hat{I}^m(\Gg)\;=\;
\lim_{N\to\infty}\,I^m_N(\Gg)
\;.
$$
%

\begin{lemma}
\label{lemma-birklikesum}
Let $g_1, g_2 \in \Qq_r(\CC)$  and $f\in C^1(S^1_\pi)$.
Furthermore let $k\geq l \geq0$ and $m\geq 1$.
Then
\begin{equation}
\label{eq-birklikesum}
\EE_\omega
\bigl(
g_1(S^{3m+k+n} \omega)\,
g_2(S^{3m+l+n}\omega)\,
f(\theta_n(\omega))
\bigr)
\;=\;
\EE\bigl(f(\theta_n)\bigr)
\,\EE(g_1\circ S^{k-l}\,g_2)
\;+\;
\Oo(m^{-\alpha})\;,
\end{equation}
uniformly in $k,l$ and $n$.
This implies,
for $\Gg(\omega,\theta)=g_1(S^{k} \omega) g_2(S^{l}\omega) f(\theta)$,
\begin{equation}
\label{eq-birklikesum1}
\hat{I}^{3m}(\Gg)
\;=\;
\EE\bigl((g_1\circ S^{k-l})\,g_2\bigr)\,I(f)
\;+\;\Oo(m^{-\alpha})
\;.
\end{equation}
\end{lemma}

\noindent {\bf Proof.}
By Lemma \ref{lemma-change-theta} and because $f$ is Lipshitz-continuous,
one has uniformly in $n$
$$
\bigl|f(\theta_n(\omega)) - 
f\left( \theta_n(\pi_{[-m-n,n+m]}(\omega))\right) \bigr| 
\;\leq\; \Oo(\lambda^\eta r^m)\;.
$$
As $g_1$ and $g_2$ are quasi-local and therefore bounded, one also deduces
uniformly in $k$, $n$ and $l$
$$
\Bigl| g_1(S^{k+n+3m}\omega)\,g_2(S^{l+n+3m}\omega) \,-\, 
\left((g_1\circ S^{k})(g_2\circ S^l)\right)\circ S^{n+3m}
\circ\pi_{[n+2m,n+k+4m]} (\omega) \Bigr|
\;\leq\;
\Oo(r^m)\;.
$$
Let us denote the two functions inside the modulus by $g$
and $\hat{g}$ respectively. 
Similarly denote $f\circ\theta_n\circ \pi_{[-n-m,n+m]}$ by $\hat{f}$.
Now consider
$\EE(g\,(f\circ\theta_n))$.
As the functions $f$ and $g$ are 
bounded, it follows from the estimates above and \eqref{eq-expmixing}
that with errors of order 
$\Oo=\Oo(m^{-\alpha})\geq\Oo(r^m)\geq\Oo(\lambda^\eta r^m)$ 
(for big $m$ and small $\lambda$)
in each step we get
$$
\EE\bigl(g\,f(\theta_n)\bigr)
\;=\;
\EE\bigl(\hat{g}\,f(\theta_n)\bigl)\,+\,\Oo
\;=\;
\EE\bigl(\hat{g}\,\hat{f}\,\bigr)\,+\,\Oo
\;=\;
\EE(\hat{g})\,\EE(\hat{f})\,+\,\Oo
\;=\;
\EE(g)\,
\EE\bigl(f(\theta_n)\bigr)\,+\,\Oo\;.
$$
This finishes the proof.
\hfill $\Box$

\vspace{.2cm}

Replacing $g_2(S^{3m+n+l}\omega)$ by $g_2(S^{l+n} \omega)$ for
$0\leq l \leq k$,
one can modify the argument by grouping 
$g_2$ and $f$ together. This gives the following

\begin{lemma}
\label{lemma-weakcorrel}
Let $g_1, g_2 \in \Qq_r(\CC)$ and let $f\in C^1(S^1_\pi)$. Then one has
for $0\leq l \leq k$ and $m\geq 1$
$$
\EE_\omega
\bigl(g_1(S^{3m+n+k}\omega)\, g_2(S^{l+n}\omega)\,f(\theta_n(\omega)\bigr)
\;=\;
\EE(g_1)\,
\EE_\omega\bigl(g_2(S^{l+n}(\omega))\,f(\theta_n(\omega)\bigr)\,+\,
\Oo(m^{-\alpha})\;,
$$
uniformly in $l,k$ and $n$. This implies for 
$\Gg(\omega,\theta)=g_1(S^{3m+k}\omega) g_2(S^l \omega) f(\theta)$
\begin{equation}
\label{eq-weakcorrel1}
\hat{I}(\Gg)
\;=\;
\EE(g_1)\,\hat{I}\bigl(g_2(S^l\omega) f(\theta)\bigr)
\,+\,\Oo(m^{-\alpha})\;,
\end{equation}
and leads, for $f=1$ and $l=0$, to 
\begin{equation}
\label{eq-weakcorrel2}
\EE\bigl(g_1(S^{3m+{k}}(\omega))g_2(\omega)\bigl)
\;=\;
\EE(g_1)\,\EE(g_2)\;+\;\Oo(m^{-\alpha})\;.
\end{equation}
\end{lemma}

\vspace{.2cm}

\noindent {\bf Proof} of Proposition~\ref{prop-birklike-birk}.
By Taylor expansions and 
$p_{1,S^{n+j}\omega}(\theta_{n+j})=p_{1,S^{n+j}\omega}(\theta_{n})+\Oo(j\lambda^\eta)$, 
one finds
$$
f(\theta_{n+6m})
\;=\;
f(\theta_n)\,+\, \lambda^{\eta}\sum_{j=0}^{6m-1} 
p_{1,S^{n+j}\omega} (\theta_n)
f'(\theta_n)\,+\, \Oo(m^2\lambda^{2\eta})\;.
$$
Therefore multiplying with $g\circ S^{6m+n}$ and averaging over $\omega$ 
and $n$ gives
$$
\hat{I}(\Gg)
\;=\;
\hat{I}^{6m}(\Gg)\,+\, \lambda^{\eta}\sum_{j=0}^{6m-1} 
\hat{I}(\Gg_{j})\,+\,\Oo(m^2\lambda^{2\eta})\,,
$$
where $\Gg_{j}(\omega,\theta)= 
g(S^{6m}\omega)p_{1,S^j\omega}(\theta) f'(\theta)$.
As $p_{1,\omega}(\theta)$ is a trigonometric polynomial in $\theta$,
Lemma~\ref{lemma-birklikesum} can be applied to each summand in order
to obtain
$$
\hat{I}^{6m} (\Gg)
\;=\;
\EE(g) I(f)\,+\,\Oo(m^{-\alpha})
\;.
$$
Because the functions $\Gg_{j}$ are uniformly bounded,
one has $\lambda^{\eta}\sum_{j=0}^{6m-1} \hat{I}(\Gg_{j})=
\Oo(m\,\lambda^{\eta})$.
Using $m=\lambda^{-\eta\frac{1}{1+\alpha}} $
now proves the first part.

\vspace{.2cm}

Now let $\EE(g)=0$. 
Again because $p_{1,\omega}$ is a trigonometric polynomial,
Lemma \ref{lemma-birklikesum} gives,
for $j\geq 3m$ and $f_j$ as defined in \eqref{eq-birklike-birk1},
$$
\hat{I}(\Gg_{j})
\;=\; 
I\left(\EE_{\omega} \big(g(S^{6m-j}\omega)p_{1,\omega}\big)f'\right)
\,+\, \Oo(m^{-\alpha})
\;=\;
I(f_{6m-j})\,+\,\Oo(m^{-\alpha})
\;.
$$
Using Lemma~\ref{lemma-weakcorrel}, one obtains for $j<3m$
$$
\hat{I}(\Gg_{j})
\;=\;
\EE(g) \hat{I}^j(p_{1,\omega}\, f'(\theta))\,+\, \Oo(m^{-\alpha})
\;=\; \Oo(m^{-\alpha})\,.
$$
All together, one has 
$$
\hat{I}(\Gg)
\;=\; 
\lambda^{\eta} 
\sum_{j=3m}^{6m-1} I(f_{6m-j}) \,+\, 
\Oo(m^2\lambda^{2\eta}, \lambda^{\eta}m^{1-\alpha}, m^{-\alpha})
\;.
$$
Because \eqref{eq-weakcorrel2} gives
$$ 
\sum_{j=3m+1}^\infty |f_{j}(\theta)|
\;=\;
\sum_{j=3m+1}^\infty 
\bigl| \EE_\omega\bigl(g(S^j\omega)p_{1,\omega}(\theta)\bigr)\,
f'(\theta)\bigr|
\;\leq\; 
C\,\sum_{j=3m+1}^\infty j^{-\alpha}
\;=\;\Oo(m^{1-\alpha})\;,
$$
one therefore deduces
$$
\hat{I}(\Gg)
\;=\; 
\lambda^{\eta} \sum_{j=1}^{\infty} I(f_{j}) \,+\, 
\Oo(m^2\lambda^{2\eta}, \lambda^{\eta}
m^{1-\alpha}, m^{-\alpha})\,.
$$
Finally choosing 
$m=\lambda^{-\frac{2 \eta}{\alpha+2}} $
concludes the proof. 
\hfill $\Box$

\section{Oscillatory sums away from band center and edges}
\label{sec-CS}

As already explained in Section~\ref{sec-dynamics}, for the calculation of the
Lyapunov exponent one needs
to evaluate the Birkhoff-like sums of functions of the type
$\Gg(\omega,\theta)=g(\omega) e^{2\imath j\theta}$, $j=1,2$. This is done
in Proposition~\ref{prop-PF} below
for energies away from the band center and band edge. By applying
it to the terms appearing when 
\eqref{eq-expan1} is replaced in \eqref{eq-lyap},
this result allows to
complete the proof of formula \eqref{eq-Thouless}. As the straight-forward
algebraic calculations are carried out in detail {\it e.g.} in \cite{CS,JSS}
and we present a similar calculation for the band edge in
Section~\ref{sec-application}, we skip the details.

\begin{proposi}
\label{prop-PF}
Let $\alpha>2$.
Suppose that the lowest order rotation phase $k$ in 
the dynamics {\rm \eqref{eq-polydef}} satisfies $d(k)=\,$
{\rm dist}$(k\!\!\mod\!
\frac{\pi}{2},0)>0$. Consider
$\Gg_j(\omega,\theta)=g(\omega) e^{2\imath j\theta}$ with $j=1,2$ and 
$g\in\Qq_r(\RR)$.
Then 
$$
\hat{I}(\Gg_j)
\;=\;
\Oo\Bigl(
\frac{\lambda^{\eta\frac{\alpha}{1+\alpha}}}{d(k)}\Bigr)
\;.
$$ 
If, moreover, $\EE(g)=0$, 
\begin{equation}
\label{eq-birklike-eval}
\hat{I}(\Gg_1)
\;=\;\,
\lambda^{\eta}\sum_{j=1}^\infty 
\EE_\omega
\bigl(\,g(S^j \omega) \overline{\beta_{1,\omega}}\,\bigr)
\,+\, 
\Oo\Bigl(
\frac{\lambda^{\eta\frac{2\alpha}{2+\alpha}}}{d(k)}\Bigr)\;.
\end{equation}
\end{proposi}

\noindent {\bf Proof.} \cite{PF,CS,JSS} 
The dynamics and the definition of the Birkhoff sums
implies $I_N(e^{2\imath j\theta})=e^{2\imath jk}I_N(e^{2\imath j\theta})
+\Oo(N^{-1},\lambda^\eta)$. This implies $I(e^{2\imath j\theta})=
\Oo(d(k)^{-1}\lambda^\eta)$. The bound
\eqref{eq-birklike-birk} of Proposition~\ref{prop-birklike-birk} thus implies
the first statement. The formula \eqref{eq-birklike-eval} now follows after a
short calculation from 
\eqref{eq-birklike-birk1}, the identity
$p_{1,\omega}(\theta)=\Im m(
\alpha_{1,\omega}-\beta_{1,\omega}e^{2\imath\theta})$ and the first statement.
\hfill $\Box$

\section{Fokker-Planck operator for drift-diffusion}
\label{sec-2ndorder-operator}

We now focus on energies for which the rotation angle $k$ in
\eqref{eq-polydef} satisfies
$k\!\!\mod\!\frac{\pi}{2}=0$ so that the argument of Proposition~\ref{prop-PF}
does not
apply in order to calculate the Birkhoff sum $I(e^{2\imath\theta})$. 
For this purpose, let us introduce the bilinear form 
$$
\langle\,g_1, g_2\, \rangle_\Omega
\;=\;
\EE_\omega\left(g_1(\omega)g_2(\omega)\right)\,+\,
2\,\sum_{m=1}^\infty 
\EE_\omega\left( g_1(\omega) g_2(S^n \omega) \right)
\;,
\qquad
g_1, g_2 \in \Qq^0_r(\RR)\;,
$$
which by \eqref{eq-weakcorrel2} is well-defined. Note that
$D_V(0)=\langle\,V, V\, \rangle_\Omega$.
Let us use the notation $p_j(\omega,\theta)=p_{j,\omega}(\theta)$ and 
$p'_j=\partial_\theta p_j$.
Then expressions like $\langle \, p_1,p'_1\,\rangle_\Omega$
are functions of $\theta$ on $S^1_\pi$.

\begin{proposi}
\label{prop-2ndorder}
Let the family $T_{\lambda,\omega}$ be as in \eqref{eq-expan},
and $F \in C^3(S^1_\pi)$. For $f\in  C^1(S^1_\pi)$ given by
\begin{equation}
\label{eq-diff-f-F}
f
\;=\; 
\langle\,p_1, p_1\,\rangle_\Omega\, F''\,+\, 
\left( \langle\,p_1, p'_1\,\rangle_\Omega \,+\, 2\,\EE(p_{2,\omega})\right)
F'
\;,
\end{equation}
one then has for $\alpha>2$
$$
I(f)
\;=\;
\Oo\bigl(\lambda^{\eta\frac{\alpha-2}{\alpha+2}}\bigr)\;.
$$
\end{proposi}

\noindent {\bf Proof.}
By a Taylor expansion, one has with errors of order 
$\Oo=\Oo(\lambda^{3\eta})$
$$
F(\Ss_{\lambda,\omega}(\theta))
\;=\; 
F(\theta) \,+\,
\sum_{k=1}^{2} \lambda^{k\eta}
p_{k,\omega}(\theta)F'(\theta)
\,+\,
\lambda^{2\eta} \frac{1}{2} 
\left[
F'(\theta) p_{1,\omega}(\theta)
p'_{1,\omega} (\theta)+
p^2_{1,\omega}(\theta)F''(\theta)
\right]\,+\,\Oo\;.
$$
We now use this for $\theta=\theta_n$ and average over $n$.
Because $p_{1,\omega}$ is centered and a polynomial,
one can apply equation \eqref{eq-birklike-birk1} 
of Proposition~\ref{prop-birklike-birk} 
to the term with power $\lambda^\eta$ and \eqref{eq-birklike-birk}
to the other terms. This gives
$$
I(F)
\;=\;
I(F) \,+\,\frac{1}{2}\,
\lambda^{2\eta}\,\Bigl( I\bigl(\langle\,p_1, p_1'\,\rangle_\Omega\,F'\bigr)
\,+\, 
I\bigl(\langle\,p_1, p_1 \,\rangle_\Omega\,F''\bigr)
\,+\, 2\,I\bigl(\EE_{\omega}(p_{2,\omega})\,F'\bigr)\Bigl)
\,+\,\Oo
$$
with errors of order $\Oo=\Oo(\lambda^{\eta\frac{3\alpha+2}{\alpha+2}})$.
As the functional $I$ is linear, resolving this equation for $I(f)$
gives the desired result.
\hfill $\Box$

\vspace{.2cm}

This proposition shows, that we can control error terms on Brikhoff 
sums for a function $f$, if $f$ is in the image of the 
operator $\Ll$ on functions on $S^1_\pi$ given by
\begin{equation}
\label{eq-backwardoperator}
\Ll
\;=\; (p\partial_\theta\,+\,q)\,\partial_\theta\;,
\qquad
p
\;=\;
\langle\,p_1, p_1 \,\rangle_\Omega\;,
\qquad
q
\;=\;
\langle\,p_1,p_1'\,\rangle_\Omega\,+\,2\,\EE(p_{2,\omega})\;.
\end{equation}
As one needs to calculate Birkhoff sums $I(f)$ pertubatively,
we are looking for some class of functions where 
$\lim_{\lambda\to 0} I(f)$ exists. For $f$ in the image under $\Ll$ of 
$C^3(S^1_\pi)$, this
limit is $0$. Thus, if this map is given by the scalar product
with some $L^2$-function $\rho$, one has
$\rho \in {\rm Ran}(\Ll)^\perp = {\rm Ker}(\Ll^*)$, where
the formal adjoint is given by
$$
\Ll^*
\;=\;
\partial_\theta(\partial_\theta p\,-\,q)\;.
$$
$\Ll^*$ is a forward Kolmogorov or Fokker-Planck operator describing the 
drift-diffusion dynamics of the process $\theta_n$ on $S^1_\pi$, and
$\Ll$ is the associated backward Kolmogorov operator \cite{Ris}. 
It will be shown that in the situations considered here,
${\rm Ker}(\Ll^*)$ is spanned
by a smooth, $L^1$-normalized function $\rho$. Furthermore,
the following theorem shows that
$f\in{\rm Ker}(\Ll^*)^\perp \cap\, C^2(S^1_\pi)$ turns out to be
sufficient for finding a solution $F\in C^3(S^1_\pi)$ of the
differential equation \eqref{eq-diff-f-F}
so that Proposition~\ref{prop-2ndorder} actually applies. 
Even though contained in \cite{SS}, let us give the proof for sake of
completeness. 

\begin{theo}
\label{theo-FP}
Suppose that $p(\hat\theta)=0$ for at most one angle $\hat\theta\in S^1_\pi$.
Furthermore suppose $q(\hat\theta)\neq 0$ in that case.
Then the Fokker-Planck operator $\Ll^*$ has a unique groundstate 
$\rho\in C^\infty(S^1_\pi)$, 
which is non-negative and normalized.
Furthermore, for $f\in C^2(S^1_\pi)$, one has
$$
I(f)
\;=\;
\int_0^\pi {\rm d}\theta\,\rho(\theta)\,f(\theta)
\;+\;
\Oo\bigl(\lambda^{\eta\frac{\alpha-2}{\alpha+2}}\bigr)\;.
$$
\end{theo}

\noindent {\bf Proof.}
Integrating the equation $\Ll^* \rho = 0$ once gives
\begin{equation}
\label{eq-dglrho}
\bigl(p\,\partial_\theta\;+\;(\partial_\theta\,p)
\;-\;q\;\bigr)\,\rho
\;=\;
C\;,
\end{equation}
where $C$ is some real constant.
As $I(f+c)=c+I(f)$ for $c=\langle\rho,f\rangle$, we may assume
$\int_0^\pi {\rm d}\theta\,\rho(\theta)\,f(\theta)\,=\,0$
once we found the normalized solution of \eqref{eq-dglrho}.
Proposition~\ref{prop-2ndorder} then gives the bound on $I(f)$ 
if one finds a solution $G\in C^2(S^1_\pi)$ of
\begin{equation}
\label{eq-dglG}
(p\partial_\theta\,+\,q)G
\;=\;
f\;,
\qquad
\int_0^\pi {\rm d}\theta\,G(\theta)
\;=\;
0\;.
\end{equation}
First let us consider the case $p>0$. Then there is no singularity and
$\Ll^*$ is elliptic. The groundstate $\rho$ 
and the function $G$ can be calculated.
For some $\tilde\theta$ set
\begin{equation}
\label{eq-wW}
w(\theta)
\;=\;
\int_{\tilde\theta}^\theta {\rm d}\xi\,\frac{q(\xi)}{p(\xi)}\;,
\qquad
W(\theta)
\;=\;
\int_{\tilde\theta}^\theta
{\rm d}\xi\;\frac{e^{w(\xi)}}{p(\xi)}\;f(\xi)\;,
\qquad
\tilde{W}(\theta)
\;=\;
\int_{\tilde\theta}^\theta
{\rm d}\xi\,e^{-w(\xi)}\;.
\end{equation}
Then
\begin{equation}
\rho
\;=\;
C_1\,\frac{e^{w}}{p}\,\bigl(C_2\,\tilde{W}\,+\,1\bigr)\;,
\qquad
G
\;=\;
e^{-w}\,\bigl(W\,+\,C_3\bigr)
\;,
\end{equation}
where $C_2$ is fixed by the condition that $\rho$ is $\pi$-periodic and
$C_1>0$ is a normalization constant.  This fixes $C=C_1\,C_2$ in
\eqref{eq-dglrho}.  
$G$ is a solution of the first equation of \eqref{eq-dglG} and for
$C\neq 0$ the constant $C_3$ is fixed by the condition that $G$ is
$\pi$-periodic. Furthermore one has
\begin{equation}
\label{eq-2condG}
0
\;=\;
\int\,\rho\,f
\;=\;
\int\,\rho\,\bigl(p\,\partial_\theta\,+\,q\bigr)G
\;=\;
-\,\int\,G\;\bigl(\partial_\theta\, p\,-\,q\bigr)\rho
\;=\;
-\,C\,
\int\,G(\theta)
\;.
\end{equation}
Thus $G$ is a solution of \eqref{eq-dglG}.
If $C=0 \Leftrightarrow C_2=0$, then $w$ is $\pi$-periodic as well as
$W$ which follows from $\int \rho f = 0$. Therefore
$G$ is $\pi$-periodic and $C_3$ is chosen such that the integral in
\eqref{eq-dglG} vanishes.

\vspace{.2cm}

Now let $p(\hat\theta)=0$ for exactly one $\hat\theta\in S^1_\pi$
and for sake of concreteness let $q(\hat\theta)> 0$ which implies
$\tilde{q}(\hat\theta)> 0$.
Then choose $\tilde\theta \in (\hat\theta,\hat\theta+\pi)$ in the first
equation of \eqref{eq-wW}, $\tilde\theta=\hat\theta$ 
in the second one and $\tilde\theta=\hat\theta+\pi$ in the third one.
As $\lim_{\theta\downarrow\hat\theta} e^{{w}(\theta)}\,=\,0$
and $\lim_{\theta\uparrow\hat\theta+\pi} e^{{w}(\theta)}\,=\,\infty$
in this case, $w$, $W$ and $\tilde{W}$ are well-defined for
$\theta\in(\hat\theta,\hat\theta+\pi)$. 
Using de l'Hospital's rule, one can prove by induction (see \cite{SS} for
details) that
$$
\rho
\;=\;
C\,\frac{e^{w}}{p}\,\tilde{W}\;,
\qquad
G
\;=\;
e^{-w}\,W
\;,
$$
can both be continued to a smooth (even at $\hat\theta$) and
$\pi$-periodic function. $C>0$ is again a normalization constant and hence
equation \eqref{eq-2condG} shows that $G$ solves \eqref{eq-dglG}.
\hfill $\Box$

\vspace{.2cm}

Before applying this result in order to prove Theorem~\ref{theo-result}, let
us present another derivation of the equation $\Ll^* \rho = 0$, albeit a
formal one, which shows that $\rho$ is
the lowest order approximation for the
assymptotic invariant measure of the process $\theta_n$.
Expanding the function $\Ss^N_{\lambda,\omega}=
\Ss_{\lambda,S^{N-1}\omega}\circ \ldots \circ \Ss_{\lambda,S\omega}
\circ\Ss_{\lambda,\omega}$ shows that the coefficients of
$$
\Ss^N_{\lambda,\omega}(\theta)
\;=\;
\theta\,+\,\lambda^{\eta}\,\hat{p}^N_{\omega}(\theta)\,+\,
\frac{1}{2}\,\lambda^{2\eta}\,\hat{q}^N_\omega(\theta)
\,+\,\Oo(\lambda^{3\eta})\;,
$$
are
$$
\hat{p}^N_{\omega}
\;=\;
\sum_{n=0}^{N-1} p_{1,S^n\omega}\;,
\qquad
\hat{q}^N_\omega
\;=\;
\sum_{n=0}^{N-1}\left(
p_{1,S^n\omega}\,+\,\sum_{j=0}^{n-1} p_{1,S^j\omega} \right)
p'_{1,S^n\omega}
\,+\,2\,\sum_{n=0}^{N-1} p_{2,S^n\omega}\;.
$$
An invariant measure $\nu_{\lambda,N}$ for $N$ steps of the dynamics $\theta_n$
on $S^1_\pi$ satisfies
\begin{equation}
\label{eq-invmsrNsteps}
\int_0^\pi \nu_{\lambda,N}({\rm d}\theta)\,f(\theta)
\;=\;
\EE\,\int_0^\pi \nu_{\lambda,N}({\rm d}\theta)\,
f\bigl(\Ss^N_{\lambda,S^{N-1}\omega}(\theta)\bigr)\;,
\qquad f \in C(S^1_\pi)\;.
\end{equation}
Supposing $\nu_{\lambda,N}({\rm d}\theta)=\rho_{\lambda,N}(\theta)\,{\rm
  d}\theta 
=\rho_N(\theta)\,{\rm d}\theta+o(\lambda^0)$,
\eqref{eq-invmsrNsteps} leads to
%
%
%
$$
\Ll^*_N\,\rho_N
\;=\;
0\,,\qquad
\Ll^*_N
\;=\;
\partial_\theta
\left(
\partial_\theta\,\EE\bigl((\hat{p}^N_{1,\omega})^2\bigr)
 -\,\EE\bigl(\hat{q}^N_\omega\bigr) 
\right)\;.
$$
Using the stationarity of $\PP$ and the definitions of 
$\hat{p}^N_{\omega}$ and $\hat{q}^N_\omega$, one deduces
$$ 
\lim_{N\to\infty}\,\frac{1}{N}\,\EE\bigl((\hat{p}^N_{1,\omega})^2\bigr)
\;=\;
p\;,\qquad 
\lim_{N\to\infty}\,\frac{1}{N}\,\EE\bigl(\hat{q}^N_\omega\bigr)
\;=\;
q\;,
$$
where the convergences are uniform in $\theta$. 
This shows that
$\frac{1}{N}\Ll_N^*\,\to\,\Ll^*$ weakly for $N\to\infty$.

\section{Application to the band center and band edge}
\label{sec-application}

This section contains the proof of Theorem~\ref{theo-result}.
Let us first consider item (i), that is the band center.
As described in Section~\ref{sec-normalform} we have to work with the 
probability space 
$\tilde{\Omega}=(\Sigma\times\Sigma)^\ZZ$ which
is isomorphic to $\Omega$ by the pairing isomorphism $\Pp$.
Using this isomorphism and the potential $V$, which is defined on $\Omega$,
let us define the two random variables on $\tilde{\Omega}$
$$
v_{\tilde\omega}
\;=\;
V(\Pp^{-1}(\tilde\omega))
\;=\;
V(\omega)\;,
\qquad
u_{\tilde\omega}
\;=\;
V(S\Pp^{-1}(\tilde\omega))
\;=\;
V(S\omega)\;.
$$
Then according to equation \eqref{eq-normform-bandcenter}
the family of matrices we have to consider is given by
$$
T_{\lambda,\tilde\omega}
\;=\;
-\,\exp\left[ \lambda\,\left(\begin{matrix}
0 & -u_{\tilde\omega} \\ v_{\tilde\omega} & 0
\end{matrix} \right)\,+\,
\frac{\lambda^2}{2}\,\left(\begin{matrix}
-u_{\tilde\omega}v_{\tilde\omega} & 2\epsilon \\
-2\epsilon & u_{\tilde\omega}v_{\tilde\omega} 
\end{matrix}\right)\,+\,\Oo(\lambda^3) 
\right]\;.
$$
In this situation one has $\alpha_{1,\tilde\omega}=\imath
(v_{\tilde\omega}+u_{\tilde\omega}) / 2\,,\,
\beta_{1,\tilde\omega}=\imath(u_{\tilde\omega}-v_{\tilde\omega})/2\,,\,
\alpha_{2,\tilde\omega}=-\imath\epsilon$ and
$\beta_{2,\tilde\omega}=-\frac{1}{2} u_{\tilde\omega}v_{\tilde\omega} $.
Using Lemma~\ref{lem-coefficients} and 
$\langle\,v-u, v-u\,\rangle_{\tilde\Omega} =2\, D_V(\pi)$ and
$\langle\,v+u, v+u\,\rangle_{\tilde\Omega} =2\, D_V(0) $,
one obtains that the polynomials \eqref{eq-backwardoperator} are 
explicitly given by
$$
p(\theta)
\;=\;
\frac{1}{2}  \,D_V(0)
\,+\, \frac{1}{2}\,D_V(\pi)\,\cos^2(2\theta)
\;,\qquad
q(\theta)
\;=\;
- \frac{1}{2}\,D_V(\pi)\, \sin(4\theta)\,-\, \epsilon\;.
$$
By assumption on $V$, one has $p>0$ uniformly on $S^1_\pi$.  
By Theorem~\ref{theo-FP} there is thus a smooth, positive and 
$L^1$-normalized groundstate $\rho_\epsilon$ 
for the operator $\Ll^*$ (which can
readily be written out).
Furthermore, one checks 
$\gamma_{1,\tilde\omega}=(v^2_{\tilde\omega}-u^2_{\tilde\omega})/2$. Then
equation  \eqref{eq-expan1}, Theorem~\ref{theo-FP} and 
Proposition~\ref{prop-birklike-birk} combined with some algebra leads to
\eqref{eq-resultbandcenter} for $\gamma_\lambda(\epsilon\lambda^2)
=\frac{1}{2}\, \gamma(\lambda)$.

\vspace{.2cm}

Now let us prove Theorem~\ref{theo-result}(ii).
Hence let 
$T_{\lambda,\omega}=N_{\lambda}
N\Tt^{-2+\epsilon\lambda^2}_{\lambda,\omega}N^{-1}
N_{\lambda}^{-1}$
be the anomaly given in \eqref{eq-bandedgeanomaly}.
As $\alpha_{1,\omega}=\imath V(\omega)/2\,,\,
\beta_{1,\omega}=-\imath V(\omega) / 2\,,\,
\alpha_{2,\omega}=-\imath(\epsilon+1)/2\,,\,$ and
$\beta_{2,\omega}=\imath(\epsilon-1)/2$, one deduces, 
using $\langle V,V\rangle_\Omega = D_V(0)$,
$$
p(\theta)
\;=\;
D_V(0) \;\cos^4(\theta)\;,
\qquad
q(\theta)
\;=\;
-\epsilon-1\,+\,(1-\epsilon)\,\cos(2\theta)
\,-\, 2\,D_V(0) \,\cos^3(\theta)\sin(\theta)\;.
$$
By assumption on $V$ one has $p(\theta) > 0$ for 
$\theta\,\not\in\,\frac{\pi}{2}$, and as $q(\frac{\pi}{2}) =-2\neq 0$,
there is a unique groundstate $\rho_\epsilon \in C^\infty(S^1_\pi)$ by 
Theorem~\ref{theo-FP}. Explicitly, one obtains
\begin{equation}
\label{eq-rhoformula}
\rho_\epsilon(\theta)
\;=\;
C\;
\int^\theta_{-\frac{\pi}{2}} d\xi\;
\frac{\cos^2(\xi)}{\cos^6(\theta)}\,
\exp\Bigl(
\frac{2}{3D_V(0)}
\,\bigl(\tan^3(\xi)-\tan^3(\theta)+3 \epsilon\tan(\xi)-3\epsilon
\tan(\theta)\bigr) 
\Bigr)
\;,
\end{equation}
where $C$ is some normalization constant.
Furthermore, one checks
$\gamma_{1,\omega}=V(\omega)^2/2$ and hence
\eqref{eq-expan1}, 
Proposition~\ref{prop-birklike-birk} and Theorem~\ref{theo-FP} imply
\eqref{eq-resultbandedge}.

\section{Bound on the quantum dynamics}
\label{sec-bound}

As already said above, the proof of Theorem~\ref{theo-logbound} follows
exactly the proof of Theorem~1 in \cite{JS} given in Section~3 and 4 therein,
except that the proof of Lemma~4 of \cite{JS} 
has to be refined in order to deal with
strong mixing \eqref{eq-expmixing} instead of independent potential values
$V(S^n\omega)$. The conclusion of the following lemma is hence exactly the same
as of Lemma~4 of \cite{JS}, and we thereby consider the proof of
Theorem~\ref{theo-logbound} to be complete.

\vspace{.2cm}

Let us set
$U=\{E\in\CC\,|\,E_0\leq \Re e (E)\leq E_1\,,\, |\Im m
(E)|\leq 1\,\}$. Furthermore introduce the transfer matrices over several
sites:
$$
\Tt^E_{\lambda,\omega}(k,m)
\;=\;
\prod^{k-1}_{n=m}\Tt^E_{\lambda,S^n\omega}
\;,
\qquad
k>m\;,
$$
Furthermore, $\Tt^E_{\lambda,\omega}(k,m)=\big(\Tt^E_{\lambda,\omega}(m,k)\big)^{-1}$
for $k<m$ and $\Tt^E_{\lambda,\omega}(m,m)=\one$.

\begin{lemma}
\label{lem-probboundsimple2}
Let $E\in U$ and $N\in\NN$. Then there is a constant 
$\hat{C}$ such that 
the set
$$
\hat{\Omega}_N(E)
\;=\;
\left\{
\,\omega\in\Omega\;\left|\;
\max_{0\leq n \leq N}\,\|\Tt_{\lambda,\omega}^E(n,1)\|^2
\,\geq\, e^{\hat{C}\,N^{\frac{1}{2}}}
\;\right.
\right\}
$$
satisfies
$$
\PP(\hat{\Omega}_N(E))
\;\geq\;
1\,-\,e^{-\,\hat{C}\,N^{\frac{1}{2}}}
\;.
$$
\end{lemma}

\noindent {\bf Proof.} For sake of notational symplicity, we will drop the 
index $\lambda$ on the transfer matrices $\Tt_{\lambda,\omega}^E$.
Let us fix $E\in U$ and $N \in \NN$ and then 
split $N$ into $\frac{N}{N_3}$ pieces of length
$N_3=N_0+N_1+2N_2$. 
For $j=0,\ldots,\frac{N}{N_3}$,  
we consider the following events:
\begin{eqnarray*}
\Omega_j^0
&=&
\left\{
\,\omega\in\Omega\;\left|\;
\|\Tt_\omega^E(jN_3+N_0,jN_3)\|\,\leq \,e^{\frac{1}{2}\gamma_0\,N_0}
\;\right.
\right\}\;,\\
\Omega_j^1
&=&
\left\{
\,\omega\in\Omega\;\left|\;
\|\Tt_{\pi_{[jN_3-N_2,N_3j+N_0+N_2]}(\omega)}^E(jN_3+N_0,jN_3)\|\,\leq 
\,e^{\frac{2}{3}\gamma_0\,N_0}
\;\right.
\right\}\;, \\
\Omega_j^2
&=&
\left\{
\,\omega\in\Omega\;\left|\;
\|\Tt_{\omega}^E(jN_3+N_0,jN_3)\|\,\leq \,e^{\frac{3}{4}\gamma_0\,N_0}
\;\right.
\right\}
\;.
\end{eqnarray*}
First we note that uniformly in $\omega$ and for some $\gamma_1>0$
$$
\|
\Tt_{\omega}^E(n,m)\|
\;\leq\;
e^{\gamma_1\,|n-m|}
\;.
$$
Therefore the hypothesis
\eqref{eq-lyaplower} implies as in the proof of Lemma~3 of \cite{JS} that,
for $E\in U$ and $N_0\in\NN$, we have
\begin{equation}
\label{eq-help}
\PP(\Omega_j^2)
\;\leq\;
1-p_0\;<\;1
\;,
\qquad
p_0>0
\;.
\end{equation}
To shorten notations let us define 
$\pi_j=\pi_{[jN_3-N_2,jN_3+N_0+N_2]}$
and $\Tt^E_{\omega,j}=\Tt^E_{\omega}(N_3j+N_0,N_3j)$.
Using the quasi-locallity of $g(\omega)=\Tt^E_\omega$ we get
\begin{eqnarray}
\left\| \Tt^E_{\omega,j}-\Tt^E_{\pi_j(\omega),j}\right\|
&=&
\left\| \sum_{l=jN_3}^{jN_3+N_0-1}\; \left(\prod_{k=jN_3}^{l-1}
 \Tt^E_{S^k\omega}\right)
\;\left[\Tt^E_{S^l\omega} - \Tt^E_{S^l \pi_j(\omega)} \right]
\left(\prod_{k=l+1}^{jN_3+N_0-1} \Tt^E_{S^k\pi_j(\omega)}\right)\;\right\| 
\nonumber
\\
&\leq& N_0\,\big(\sup_\omega(\Tt^E_\omega)\big)^{N_0-1}\,C\,r^{N_2}
\;,\nonumber
\end{eqnarray}
where $C=C(g)$ as in \eqref{eq-quasilocal}. Now
choosing $N_2=cN_0$ for an adequate constant $c$, it follows that
$$
\| \Tt^E_{\omega,j} - \Tt^E_{\pi_j(\omega),j} \|
\;\leq\;
e^{\frac{1}{2}\gamma_o\,N_0}
$$
Therefore for $\omega\in \Omega_j^0$
$$
\|\Tt^E_{\pi_j(\omega),j}\|
\;\leq\;
\|\Tt^E_{\omega,j}\|\,+\, e^{\frac{1}{2}\gamma_0\, N_0}
\;\leq\;
2\, e^{\frac{1}{2}\gamma_0\, N_0}
\;\leq\;  e^{\frac{2}{3}\gamma_0\, N_0}
$$
for $N_0$ large enough, implying $\Omega_j^0 \subset \Omega_j^1$.
By a similar calculation, one obtains the second inclusion of
\begin{equation}
\Omega_j^0 \;\subset\; \Omega_j^1
\;\subset\; \Omega_j^2\;.
\label{eq-subsetsequence}
\end{equation}
By \eqref{eq-help} this implies
$$ 
\PP(\Omega_j^1)
\;\leq\;\PP(\Omega_j^2)
\;\leq\;1-p_0\;.
$$

\noindent Now clearly $\Omega_j^1$ is
$\pi_j=\pi_{[jN_3-N_2,jN_3+N_0+N_2)]}$-measurable. 
Therefore the strong mixing condition \eqref{eq-cylinder} implies
that $\PP(\Omega_0^1\cap\Omega_1^1)\leq
\PP(\Omega_0^1)\PP(\Omega_1^1)\,(1+CN_1^{-\alpha})
\leq (1-p_0)^2\,(1+CN_1^{-\alpha})$. 
At the next step, one obtains
$\PP(\Omega_0^1\cap\Omega_1^1\cap
\Omega_2^1)\leq (1-p_0)^3\,(1+CN_1^{-\alpha})^2$. 
Iteration and \eqref{eq-subsetsequence} therefore give
$$
\PP
\Bigl(\;
\bigcap_{j=0,\ldots,N/N_3}
\Omega_j^0
\Bigr)
\;\leq\; 
\PP
\Bigl(\;
\bigcap_{j=0,\ldots,N/N_3}
\Omega_j^1
\Bigr)
\;\leq\; 
\left(\,
(1-p_0)(1+CN_1^{-\alpha})\,\right)^{\frac{N}{N_3}}
\;.
$$
Now let us choose $N_1$ sufficiently large such that 
$1-p_1=(1-p_0)(1+CN_1^{-\alpha})<1$. Then
$$
\PP\left(
\left\{
\,\omega\in\Omega\;\left|\;
\max_{0\leq j \leq N/N_3}\,\|\Tt_\omega^E(jN_3+N_0,jN_3)\|^2
\,\leq\, e^{\gamma_0\,N_0}
\;\right.
\right\}
\right)
\;\leq\;
(1-p_1)^{\frac{N}{N_3}}\;.
$$
Furthermore
$\Tt_\omega^E(jN_3+N_0,jN_3)=
\Tt_\omega^E(jN_3+N_0,1)\Tt_\omega^E(jN_3,1)^{-1}$. 
As $A=BC$ implies either
$\|B\|\geq \|A\|^{\frac{1}{2}}$ or $\|C\|\geq \|A\|^{\frac{1}{2}}$ for
arbitrary matrices, and $\|A^{-1}\|= \,\|A\|$ for
$A\in\,$SL$(2,\CC)$, it therefore follows that
$$
\PP\left(
\left\{
\,\omega\in\Omega\;\left|\;
\max_{0\leq j \leq N/N_3}\,
\max\{
\|\Tt_\omega^E(jN_3,1)\|^2,
\|\Tt_\omega^E(jN_3+N_0,1)\|^2
\}
\,\geq\, \;e^{\frac{1}{2}\,\gamma_0\,N_0}
\;\right.
\right\}
\right)
$$
is greater or equal than $1-(1-p_1)^{\frac{N}{N_3}}$.
Choosing $N_0=cN^{\frac{1}{2}}$ with adequate $c$ concludes the proof.
\hfill $\Box$



\begin{thebibliography}{999}


\bibitem[BS]{BS} J. Bourgain, W. Schlag, {\sl Anderson Localization for
Schr\"odinger Operators on $\ZZ$ with Strongly Mixing Potentials},
Commun. Math. Phys. {\bf 215}, 143-175 (2000).

\bibitem[Bra]{Bra} R. C. Bradley, {\sl Basic Properties of Strong Mixing
Conditions. A Survey and Some Open Questions}, Probability Surveys {\bf 2}, 
107-144 (2005).  

\bibitem[Bow]{Bow} R. Bowen, {\sl Equilibrium States and Ergodic Theory of
Anosov Diffeomorphisms}, Lect. Notes in Math. {\bf 470}, (Springer, Berlin,
1975).  

\bibitem[CS]{CS}
V. Chulaevsky, T. Spencer, {\sl Positive Lyapunov exponents for a class
of deterministic potentials}, Commun. Math. Phys. {\bf 168}, 455-466 (1995)

\bibitem[Dam]{Dam} D. Damanik, {\sl Lyapunov exponents and spectral analysis
of ergodic Schr\"odinger operators: A survey of Kotani theory and its
applications}, preprint 2006, to appear in Barry Simon Festschrift. 

\bibitem[DG]{DG} B. Derrida, E. J. Gardner, 
{\sl Lyapunov exponent of the one dimensional Anderson model: weak
disorder expansion}, J. Physique {\bf 45}, 1283-1295 (1984).

\bibitem[Gou]{Gou} S. Gou\"ezel, {\sl Sharp polynomial estimates for the decay
of correlations}, Israel. J. Math. {\bf 139}, 29-65 (2004).

\bibitem[Gua]{Gua} I. Guarneri,
{\sl Spectral properties of quantum diffusion
on discrete lattices}, Europhys. Lett., {\bf 10}, 95-100, (1989); {\sl On an
estimate concerning quantum diffusion in the presence of a fractal spectrum}, 
Europhys. Lett., {\bf 21}, 729-733, (1993).

\bibitem[Jit]{Jit} S. Jitomirskaya,
{\sl Ergodic Schr\"odinger operators {\rm (}on one foot{\rm)}.} 
preprint 2006, to appear in Barry
Simon Festschrift.

\bibitem[JS]{JS} S. Jitomirskaya, H. Schulz-Baldes,
{\sl Upper bounds on wavepacket spreading for random Jacobi matrices}, 
to appear in Commun. Math. Phys. (2006).

\bibitem[JSS]{JSS} S. Jitomirskaya, H. Schulz-Baldes, G. Stolz,
{\sl Delocalization in random polymer chains}, Commun.
Math. Phys. {\bf 233}, 27-48 (2003).

\bibitem[KW]{KW} M. Kappus, F. Wegner, {\sl Anomaly in the band
centre of the one-dimensional Anderson model}, Z. Phys. {\bf B 45}, 15-21 
(1981).

\bibitem[PP]{PP} W. Parry, M. Pollicott, {\sl Zeta functions and the periodic
orbit structure of hyperbolic dynamics}, Ast\'erisque {\bf 187-188},
(Soc. Math. de France, 1990).

\bibitem[PF]{PF} L. Pastur, A. Figotin, {\sl Spectra of Random and
Almost-Periodic Operators}, (Springer, Berlin, 1992).

\bibitem[Ris]{Ris} H. Risken, {\sl The Fokker-Planck equation}, Second
Edition, (Springer, Berlin, 1988).

\bibitem[Sch]{S} H. Schulz-Baldes, {\sl Lyapunov exponents at anomalies 
of \mbox{${\rm SL}(2,\RR)$} actions}, 
to be publ. in Operator Theory, Advances and Applications, (Birkh\"auser,
Basel, 2006).

\bibitem[SS]{SS} C. Sadel, H. Schulz-Baldes, {\sl Scaling diagram for the
localization length at a band edge}, preprint, {\tt math-ph/0702051}.

\bibitem[Tho]{Tho} D. J. Thouless, in Ill-Condensed Matter, Les
Houches Summer School, 1978, edited by R. Balian, R. Maynard,
G. Toulouse (North-Holland, New York, 1979).

\end{thebibliography}
\end{document}